# Controlled Growth of Bismuth Ferrite Multiferroic Flowers

B. Andrzejewski[a*], K. Chybczyńska[a], B. Hilczer[a], M. Błaszyk[a], T. Luciński[a], M. Matczak[a,b],
L. Kępiński[c]

[a] Institute of Molecular Physics
Polish Academy of Sciences
Smoluchowskiego 17, PL-60179 Poznań, Poland
* corresponding author: bartlomiej.andrzejewski@ifmpan.poznan.pl

[b] NanoBioMedical Centre, Adam Mickiewicz University
Umultowska 85, PL-61614 Poznań, Poland

[c] Institute of Low Temperature and Structure Research
Polish Academy of Sciences
Okólna 2, PL-50422 Wrocław, Poland

*Abstract* — **This study reports on the synthesis of ball-like bismuth ferrite $BiFeO_3$ nanoflowers by means of microwave assisted hydrothermal process and also on their composition and mechanism of growth. It turns out that the petals of the nanoflowers are composed of the nanocrystals with the size about 35-39 nm whereas their thickness and size depends on the concentration of surfactants. The petals contain $BiFeO_3$ phase and traces of $Bi_2O_3$ oxide and metallic Bi and Fe deposited mainly at their surface. Amounts of impurity phases are more pronounced in nanoflowers synthesized during short time, and become almost negligible for longer microwave processing. The nanoflowers contain also mixed Fe valence, with the $Fe^{2+}/Fe^{3+}$ ratio depending on the time of synthesis. The growth and shape of the nanoflowers result from the process of diffusion in the initial stages of hydrothermal reaction.**

*Keywords-component; bismuth ferrite, multiferroics, nanoflowers, microwave assisted synthesis*

## I. INTRODUCTION

Nanoparticles exhibit a tendency to aggregate during the synthesis process in a variety of form, the most intersting of which are nanostructures very similar in form to living plants. Among them, one can mention the most simple shapes like nanoalgae [1], more complex nanodendrites [2], nanograss [3, 4] nanotrees [4] and of course the most sophisticated forms like nanoflowers [5]. Nice SEM images of nanoplants and nanoforms have been awarded by the Materials Research Society on the Science as Art competition organized twice a year [6]. However, these nanostructures are not only beautiful but also important for understanding their physical nature as well as from the point of view of their future applications. For example, nanoplants like nanoseeweeds and nanotrees when sensitized by organic dyes become effective solar cells [4, 7]. Nanograss lithium batteries are heavy-duty source of power for mobile phones [3]. Nanoflowers can be used as excellent field emitters [8, 9], they also exhibit high catalytic [10] and photocatalytic activity [11] and enhanced dielectric response [12]. Due to excellent biocompability of some nanoflowers they are also important for applications in medicine and biology as amperometric [13] and colorimetric [14] biosensors, cell tags for in-vivo applications [15], for cancer cell recognition, bioimaging [16, 17] and drug delivery [18].

The dimension of nanoflowers varies from a few dozen of nm (about 40 nm for Au nanoflowers) [15] to a few dozen of micrometers (about 50 µm for $SnO_2$ flowers) [19]. Thus the common term "nanoflowers" does not necessarily refer to the external size of these objects. It rather comprises the flower-like materials with characteristic lenghts below 100 nm in at least one dimension (for example thickness of the petals). Nanoflowers can consist of plate- or sheet-like petals [20], perforated or brush-like ones [21], nanocrystalline petals [22], petals branched into tips [10], nanobelt-like petals [23], nanofibers [12], and even of bundles of nanorods [24]. They exhibit sometimes unusual morphological details, for example hollow cores [25], vase like [26] and hexangular shapes [27] or snowflake-like [13, 28] and downy-velvet-flower-like nanostructures [29].

Nanoflowers can be made of various elements like metals [15, 30, 31], carbon, [25] and of compounds of the elements. Examples of the latter are metal oxides [9, 12, 20-23] and various salts: suphides, tellurides, nitrides and phosphides [24, 27, 29, 32-35]. There are also known organic-inorganic nanoflowers [36] and DNA nanoflowers [17].

Recently, first nanoflowers of functional materials like multiferroics have been synthesized [37, 38]. Multiferroics exhibit simultaneously more than one order parameter in a single phase and therefore have large technological potential. The most interesting are magnetoelectric (ME) multiferroics with magnetic and charge ordering and some mutual coupling between magnetization and spontaneous polarization. Bismuth ferrite $BiFeO_3$ (BFO) is the best known material which exhibit ME multiferroic properties at room temperature [39]. It belongs to rhombohedrally distorted perovskites with $R3c$ space group and has high ferroelectric Curie temperature $T_C=1100$ K and high Néel temperature $T_N=643$ K. The ferroelectric (FE) properties result from the ordering of lone electron pairs of $Bi^{3+}$, whereas the antiferromagnetic (AFM) G-type ordering of $Fe^{3+}$ spins exhibits cycloidal modulation with the period $\lambda=62$ nm [40, 41]. It is assumed that a weak FM moment in this compound originates from Dzyaloshinskii-Moriya type interaction which forces small canting of the spins out of the rotation plane of the cycloid. The weak FM moment increases when the spin cycloid is suppressed in BFO particles with sizes smaller or comparable to the modulation period $\lambda$. The spin ordering in the cycloid can be also modified by strong magnetic field [40, 42].

The paper reports on the process of growth of bismuth ferrite multiferroic flowers obtained very recently by microwave assisted hydrothermal synthesis [43]. The method to obtain high-purity BFO phase and the parameters important for controlling the growth and morphology of BFO nanoflowers are discussed. One can expect that these materials can be applied in spintronic devices, as THz radiation emitters or catalysts [44].

## II. EXPERIMENTAL

*A. Sample Synthesis*

Powder-like samples composed of BFO nanoflowers were synthesized by means of microwave assisted hydrothermal method [43]. The nitrate of bismuth $Bi(NO_3)_3 \cdot 5H_2O$ and iron $Fe(NO_3)_3 \cdot 9H_2O$ in molar ratio 1:1 were used as the precursors for the synthesis. The precursors were added together with $Na_2CO_3$ into a KOH water solution of a molar concentration of 6 M. To control the process of growth of BFO nanoflowers various amounts of polyethylene glycol PEG 2000 were added to the mixtures. The mixtures were transferred into a Teflon reactor (XP 1500, CEM Corp.), loaded into a microwave oven (MARS 5, CEM Corp.) and heated at the 200 °C during $t_s=30$ min or 60 min. After the synthesis, the suspensions of BFO nanoflowers were first cooled to room temperature, next collected by filtration kit, rinsed with $HNO_3$, distilled water and placed in a dryer for 2 h. The final products were brown powders of BFO nanoflowers.

*B. Sample Characterization*

The crystallographic structure of the $BiFeO_3$ nanoflowers were studied by means of X-ray diffraction method (XRD) using a diffractometer fitted with a Co lamp ($\lambda=0.17928$ nm) and with a HZG4 goniometer in the Bragg-Brentano geometry.

Scanning electron microscope (SEM) FEI NovaNanoSEM 650 and transmission electron microscope (TEM) Philips CM20 SuperTwin were used to study the morphology and the structure of $BiFeO_3$ nanoflowers.

X-ray photoelectron spectroscopy (XPS) was applied to study the composition of the BFO nanoflowers. The spectra were collected at room temperature with UHV (standard pressure of $5 \cdot 10^{-10}$mbar) VG Scienta R3000 spectrometer and $AlK_\alpha$ radiation (1,486.6 eV). $BiFeO_3$ in the form of fine powder was spilt onto conducting adhesive carbon tape fixed to a molybdenum support. In every case the binding energy was determined by reference to the C1*s* component at the energy of 285 eV and Gaussian-Lorentzian functions were used to deconvolute the line shapes.

Magnetic measurements were performed using a Quantum Design Physical Property Measurement System (PPMS) fitted with a Vibrating Sample Magnetometer (VSM) probe.

## III. RESULTS AND DISCUSSION

An example of the SEM micrograph of an assembly of BFO nanoflowers resulting from the microwave synthesis during $t_s=60$ min is presented in Fig. 1. The almost regular ball-like nanoflowers with diameter in the range of 6 – 21 µm are composed of fine petals.

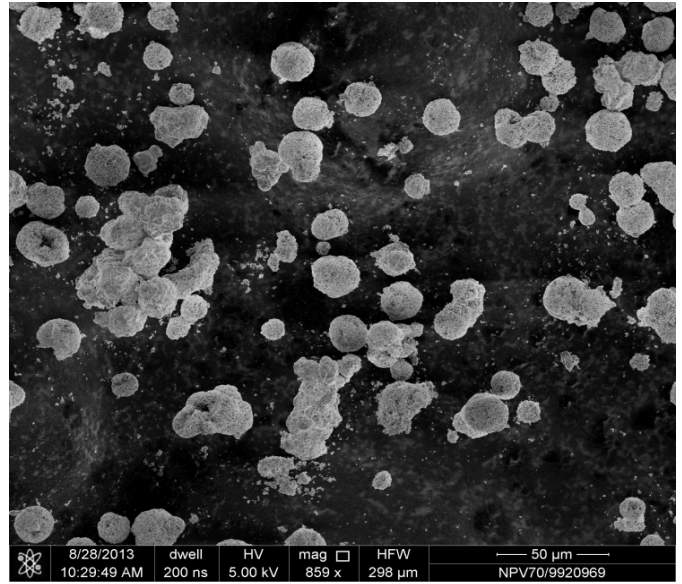

Figure 1.  BFO nanoflowers obtained during $t_s$= 60 min synthesis.

The morphological details of a selected BFO nanoflower obtained in the same synthesis conditions are shown in Fig. 2. One can observe that the ball-shaped nanoflower consist of a great number of thin crystalline petals arranged perpendicularly to the nanoflower surface.

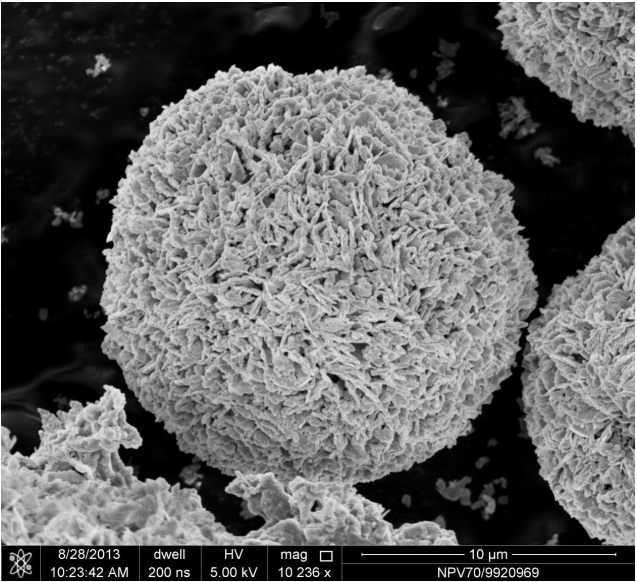

Figure 2. Morphological details of a selected BFO nanoflower; the diameter of the ball-like nanoflower amounts to ~17 μm.

We have found that the shape of BFO nanoflowers depends on the progress of microwave assisted reaction. If the reaction is in an initial stage, and the time $t_s$ of the synthesis does not exceed about the 20 min. the nanoflowers are irregular and composed of a few dozen of the petals. In the initial stage of the synthesis small nanoflower "buds" appear, as shown in Fig. 3a, which are growing and blooming (Fig. 3b and 3c) as the reaction proceeds. The nanoflowers are fully developed when the reaction is completed, which takes place after about $t_s$=60 min of processing. The developed nanoflowers form regular balls consisting of few hundreds of platelet-like petals as shown in Fig. 2 and 3d. Thus we can assume that the growth of the nanoflowers begins from only a few petals forming central part of flowers, which is very pronounced in Figs. 3a and 3b. The succeeding petals start to crystallize around these flower "buds".

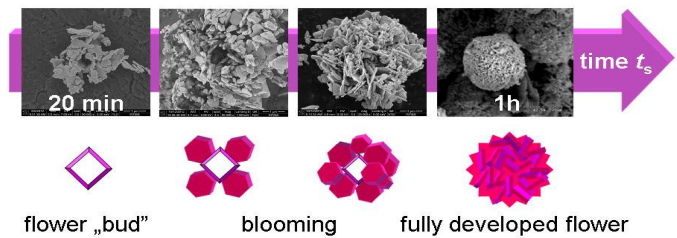

Figure 3. Growth process of BiFeO$_3$ flowers (the pictures are not in the same scale)

One can observe that the individual petals are irregular crystallites with thickness ranging from a few dozen of nanometers to about 500 nm and longitudinal dimension of few hundreds nm. The growth mechanism of such thin crystallites is determined by the processes taking place during the hydrothermal synthesis:

Bi(NO$_3$)$_3$·5H$_2$O + Fe(NO$_3$)$_3$·9H$_2$O + 6KOH ⇒ BiFeO$_3$ + 6KNO$_3$ + 17H$_2$O.

Crucial here are the intermediate stages of this reaction during which bismuth and iron oxides are formed [37]:

X(NO$_3$)$_3$ → X$^{3+}$ + 3NO$_3^-$ ↑
X$^{3+}$ + 3OH$^-$ → X(OH)$_3$
2X(OH)$_3$ → X$_2$O$_3$ + 3H$_2$O
where: X=Bi, Fe.

Bismuth ferrite compound originates due to the diffusion between Bi$_2$O$_3$ and Fe$_2$O$_3$ oxides occurring in the last stage of the reaction:

Bi$_2$O$_3$ + Fe$_2$O$_3$ → 2BiFeO$_3$

The process of diffusion strongly influences the direction of crystallographic growth resulting in formation of dendrites and finally in petal-like morphology [45]. The growth of petals can not be related to Wulff facets theorem [46, 47] because the rhombohedral compounds like BFO crystallize without any preferred growth orientation. However, it explains the ball-like agglomeration of the petals forming the flowers: the minimization of the surface to volume ratio minimizes also the high surface energy of the BFO agglomerates.

The thickness of the petals was found to be highly dependent upon the amount of PEG addition. The distribution of the petals thicknesses can be well fitted using log-normal function [48]:

$$f(D) = \frac{1}{\sqrt{2\pi\sigma^2}} \frac{1}{D} exp\left(-\frac{1}{2\sigma^2} ln^2\left(\frac{D}{D_m}\right)\right) \quad (1)$$

where $D_m$ denotes the median thickness of the petal and $\sigma$ is the distribution width. The histograms of the petals thicknesses together with the fits obtained by means of eq. (1) are shown in Fig. 4. For the low amount of PEG 2000 used during synthesis i.e. 0% and 0.01% the histograms can be fitted with single log-normal function. However, if the concentration of PEG is higher, the histograms exhibits two well resolved maxima and a superposition of two log-normal functions is necessary to obtain a reasonable fit. The parameters of the best fits are given in Table I.

Table 1 Mean size <D>, median thickness $D_m$ and distribution width $\sigma$ of the petals as dependent on the amount of PEG 2000 used in the synthesis

| wt% PEG | <$D$> [nm] | $D_{m1}$ [nm] | $\sigma_1$ | $D_{m2}$ [nm] | $\sigma_2$ |
|---|---|---|---|---|---|
| 10% | 524 | 348 | 0.2 | 642 | 0.1 |
| 1% | 260 | 138 | 0.6 | 371 | 0.1 |
| 0.2% | 252 | 130 | 0.5 | 424 | 0.1 |
| 0.01% | 159 | 120 | 0.3 | - | - |
| 0% | 78 | 61 | 0.4 | - | - |

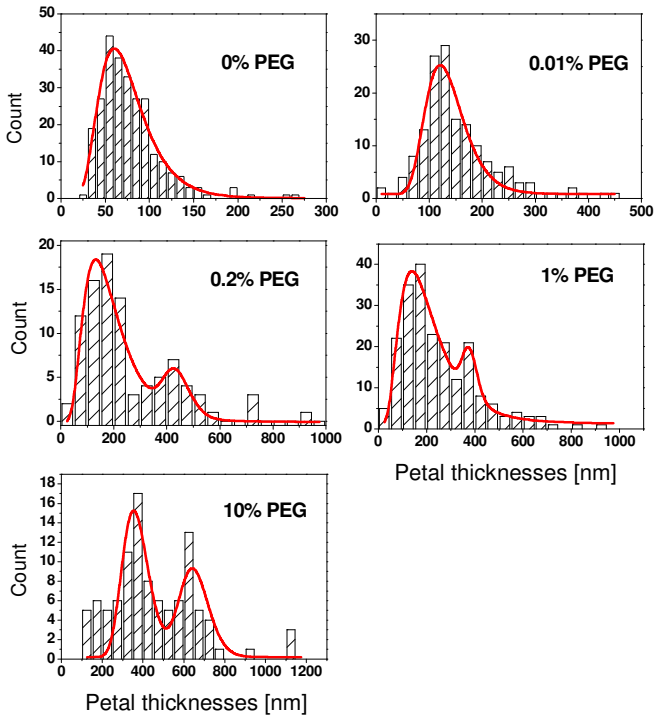

Figure 4. Histograms of the petal thicknesses of selected BFO nanoflowers.

It is evident, that the nanoflowers synthesized without PEG are composed of the finest petals. In this case, the thickness of petals varies from about 30 nm to about 200 nm, as presented in the histogram in Fig. 4a. Most of the petals exhibit thickness between 50 nm and 100 nm with the mean size $\langle D \rangle \approx 78$ nm. For higher amount of PEG 2000 addition the mean thickness of petals increases and above 0.2% of PEG used for the synthesis the histograms exhibit two maxima. The distribution of the petals thicknesses obtained for the highest amount of PEG (10%) is very wide with two maxima: at about 350 nm and 650 nm. We assume that the addition of PEG promotes two mechanisms of growth of BFO crystals in the nanoflowers. The first one is the increase in the thickness of flat crystals/petals. The second mechanism leads to the formation of very thick petals similar to microcubes. Indeed, the nanoflowers obtained for a very high concentration of PEG are very dense and composed both of the fine petals and of the cube-like BFO crystals. This is another manifestation of the fact that BFO rhombohedral compound crystallizes without any preferred orientation. We should admit that BFO microcubes can be also easily obtained by means of microwave assisted synthesis [38].

The crystalline structure of the petals was also examined by means of TEM. Results are presented in Figs. 5 and 6. The petals of nanoflowers obtained during long-time synthesis (at least $t_s$=60 min) are composed of small nanocrystals (see Fig. 5). Selected area diffraction pattern (SAD) presented in the inset to Fig. 5 indicates that these nanocrystals exhibit well crystallized bismuth ferrite phase. On the other hand, the petals of the nanoflowers obtained during a short time microwave heating ($t_s$=20 min) contain a great amount of small BFO nanocrystallites (of a few nm in size) immersed in an amorphous phase. Wide diffraction rings in the SAD pattern of this petal (inset to Fig. 6) confirm this interpretation.

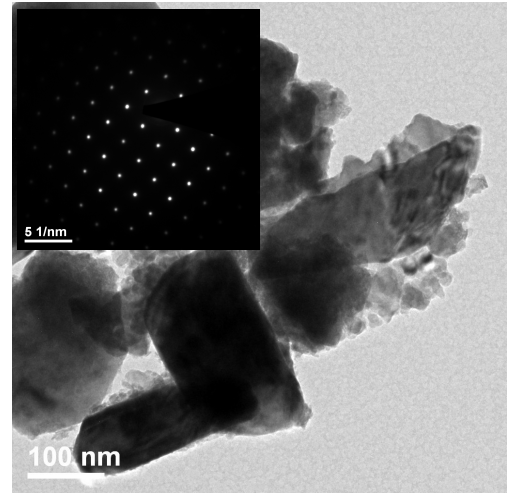

Figure 5. TEM micrograph of the biggest crystallites for the BFO nanoflowers synthesised during long time $t_s$=60 min. The inset presents SAD pattern.

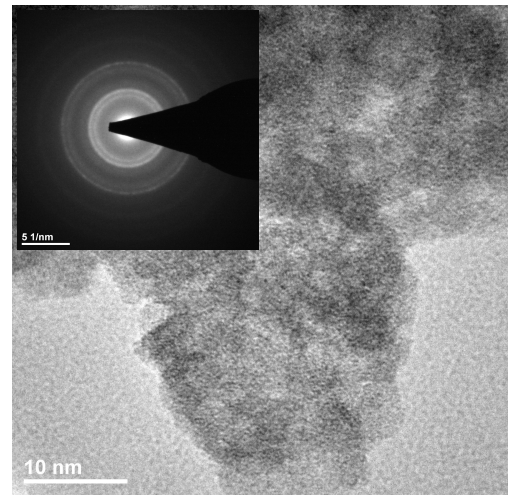

Figure 6. TEM micrograph for the BFO nanoflowes synthesised during short time $t_s$=20 min. The insert to figure shows SAD pattern.

The XRD patterns measured at room temperature for the powders containing BFO nanoflowers are presented in Fig. 7. Panel 7a shows the pattern for the powder synthesized during the long time ($t_s$=60 min). The rest of the panels present the data for the powders obtained after $t_s$=30 min of processing and with various concentration of PEG 2000 surfactant (from 0.01 to 10 wt%). The solid lines correspond to the best fits obtained by means of Rietveld method and calculated using FULLPROF software to the experimental data represented by open points. The lines below the XRD data indicate the difference between the data and the fit. The vertical sections give the positions of individual Bragg peaks. Analysis of these XRD patterns indicates the presence of the BFO rhombohedral phase with *R3c* space group in all samples. The parameters of the hexagonal crystallographic cell are given in Table II. It should be however, noted that XRD studies of nanopowders have a considerable draw-back due to a decrease in the diffraction coherence length. As a result, the decrease in the grain/crystallite size causes an increase in the halfwidth of the diffraction profile.

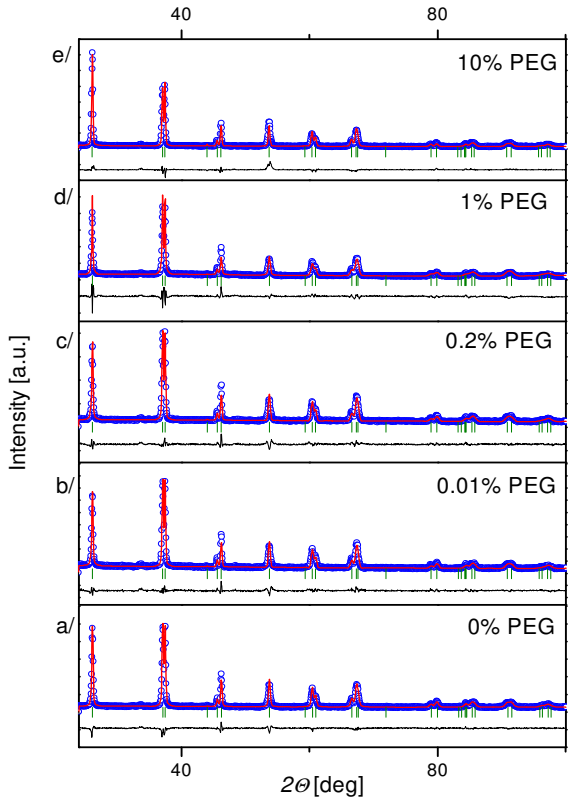

Figure 7. The XRD pattern for the powders containing BFO nanoflowers.

Table II The parameters of the hexagonal $R3c$ crystallographic cell for BFO nanoflowers synthesized during $t_s$=30 min with various PEG 2000 content

| wt % PEG | a, b [Å] | c [Å] | a,b, g [deg] | d [nm] | v [Å]$^3$ |
|---|---|---|---|---|---|
| 10% | 5.579 | 13.865 | a=b= 90°, g= 120° | 39 | 373.685 |
| 1% | 5.575 | 13.856 | a=b= 90°, g= 120° | 35 | 372.966 |
| 0.2% | 5.576 | 13.857 | a=b= 90°, g= 120° | 35 | 373.086 |
| 0.01% | 5.576 | 13.856 | a=b= 90°, g= 120° | 35 | 373.028 |
| 0% | 5.572 | 13.847 | a=b= 90°, g= 120° | 36 | 372.326 |

The mean size of the crystallites in the powders was calculated using the Scherrer's equation [49]: $d=K\lambda/\beta\cos\Theta$, where $d$ denotes the crystallite size, $\beta$ is the half-width of the diffraction peak (012), $\Theta$ stands for the angle corresponding to the position of the Bragg peak and $\lambda$ is the used wavelength. The value of the constant K in the Scherrer's equation was assumed K=0.9. Fig. 8 shows the mean size of the nanocrystallites forming the petals *versus* the PEG concentration in the solution. One can observe that the size of the nanocrystallites weakly depends on the PEG concentration. This is in clear contrast to the strong dependence of the mean thickness of the petals $D_m$ on the PEG 2000 content in the initial solution. Thus we conclude that the petals, regardless of their thickness, are composed of the same building blocks *i.e.* similar in size BFO nanocrystals.

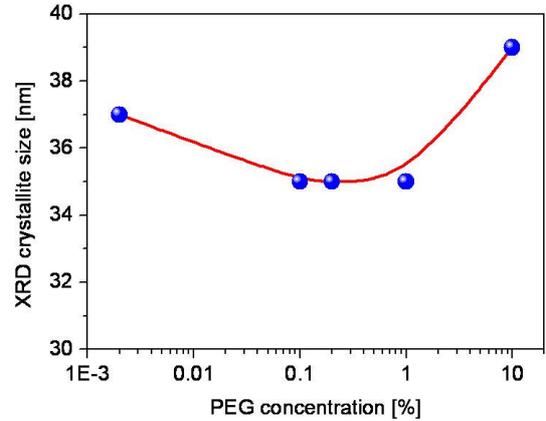

Figure 8. Dependence of the crystalline size *d* on the PEG 2000 concentration

Since the XRD pattern indicate almost no impurities, secondary or ternary phases, we performed also elemental analysis of the BFO nanoflowers by means of EDX mapping of elements distribution. The maps are presented in Fig. 9.

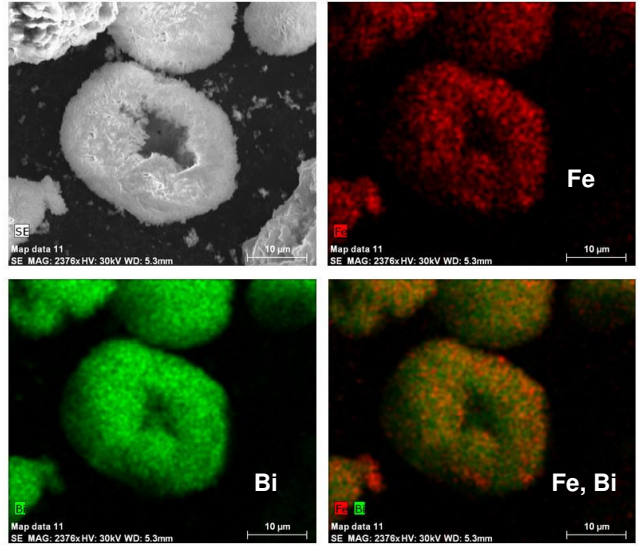

Figure 9. SEM micrograph and the distribution of the elements (Fe- red colour, Bi- green color) in BFO nanoflowers.

The distribution of Bi and Fe elements is homogeneous, even inside of the BFO nanoflower, where we expected to find impurity phases, which may serve as a "glue" for crystalline petals. More information on the composition of the nanoflowers was obtained from XPS analysis. The results are shown in Figs. 10, 11 and 12. The signal from Bi$4f$ core level in BFO with maxima corresponding to Bi$4f_{5/2}$→164.9 eV and Bi$4f_{7/2} \to$ 159.6 eV is shown in Fig. 10. The spectrum was deconvoluted to extract the contributions originating from Bi$^{3+}$ ions in BFO compound, Bi$^0$ in metallic bismuth and Bi$^{3+}$ ions composing Bi$_2$O$_3$ oxide. In the case of the BFO powder synthesized for the short time $t_s$=20 min, contributions from metallic bismuth and Bi$_2$O$_3$ oxide are substantial and equal to about 30% and to 20%, respectively. The amount of the BiFeO$_3$ phase is about 50%, only. These values, however, doesn't correspond to the bulk composition of BFO nanoflowers, because the XPS signal comes from the thin layer (about 5 nm) at the surface of the petals. Indeed, the XRD data for this sample indicate negligible amounts of impurities (see Fig. 7a). Therefore, it is reasonable to conclude that the impurities are deposited mainly at the surface of the flowers. The XPS study performed for the flowers obtained with a long-time synthesis ($t_s$=60 min) indicates traces of metallic bismuth Bi$^0$ and Bi$_2$O$_3$ oxide at the surface. However, the overlapping of the Bi$^{3+}4f$ core level doublet of BFO with traces of the doublet of metallic Bi$^0 4f$ core level and also with Bi$^{3+}4f$ core level in Bi$_2$O$_3$ hinders more detailed analysis.

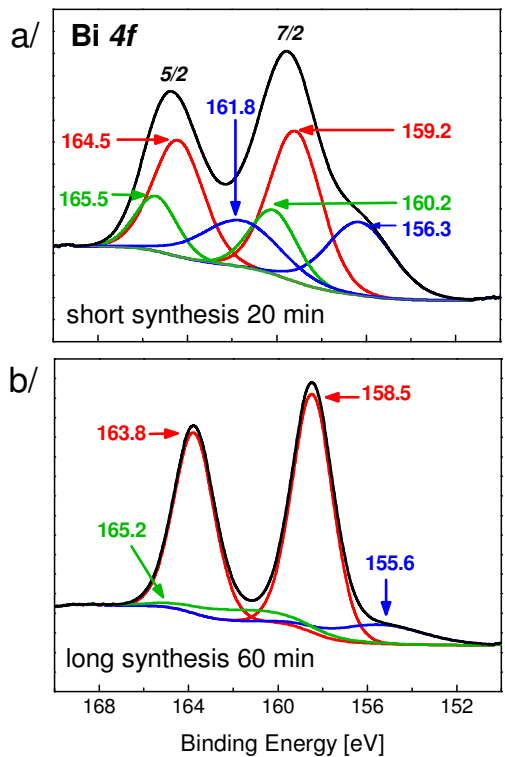

Figure 10. XPS signal from Bi$4f$ core level electrons for the sample synthesised during short time $t_s$=20 min a) and long time $t_s$=60 min b)

Bi$_2$O$_3$ oxide and metallic Bi and Fe deposited at the surface of the BFO flowers are formed during intermediate stages of BFO compound synthesis, as discussed above.

The XPS signal from Fe$2p$ core level in BFO is presented in Fig 11. The total XPS signal for the sample synthesized during short time $t_s$=20 min shown in Fig. 11a can be deconvoluted into the following set of synthetic peaks in BFO phase: Fe$2p_{1/2} \to$ 724.9 eV and Fe$2p_{3/2} \to$ 711.5 eV due to Fe$^{2+}$ ions, Fe$2p_{1/2} \to$ 727.0 eV and Fe$2p_{3/2} \to$ 713.7 eV due to Fe$^{3+}$ ions. The contribution from Fe$^0 2p_{3/2} \to$ 708.8 eV transition in metallic iron Fe$^0$ is shifted towards lower energies. The content of Fe$^{2+}$ and Fe$^{3+}$ ions in the sample synthesized during short time $t_s$=20 min (Fig. 11a) is almost identical and equal to about 51% of Fe$^{2+}$ ions and to about 49% of Fe$^{3+}$ ions.

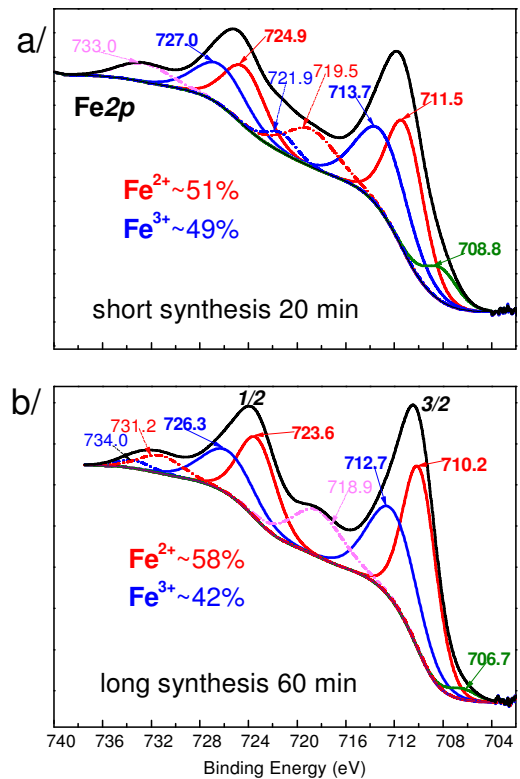

Figure 11. XPS signal from Fe$2p$ core level electrons for the sample synthesized during short time $t_s$=20 min a) and long time $t_s$=60 min b)

The XPS signal recorded for the nanoflower processed during long time $t_s$=60 min is presented in Fig. 11b. The main contributions to the total signal originate from BiFeO$_3$ compound expressed as two synthetic peaks due to Fe$^{2+}$ ions: Fe$2p_{1/2}$ at 723.6 eV and Fe$2p_{3/2}$ at 710.2 eV; and the other two synthetic peaks corresponding to Fe$^{3+}$ ions: Fe$2p_{1/2} \to$ 726.3 eV, Fe$2p_{3/2} \to$ 712.7 eV. The signals from Fe$^{2+}$ and Fe$^{3+}$ ions are overlapped due to strong multiplet splitting and shake up phenomena (satellites marked in Fig. 11b by „dash-dot"). The weak signal caused by metallic iron presence (Fe$^0$) at 706.7 eV is shifted towards lower energies like in the sample processed for the short time $t_s$=20 min. The content of Fe$^{2+}$ and Fe$^{3+}$ ions is this time unequal and is about 58% and 42%, respectively. The differences in the ratio between Fe$^{2+}$ and Fe$^{3+}$ ions can be caused by the reaction between metallic Fe and Bi$_2$O$_3$ deposited on the surface of petals. The product of this reaction BiFeO$_x$ should exhibit oxygen deficiency and thus higher ratio of Fe$^{2+}$ ions. Simultaneously, the content of metallic Fe and Bi$_2$O$_3$ impurities should decrease as it is really observed in XPS study for the nanoflowers synthesized during long time.

The O*1s* binding energy of the BiFeO$_3$ phase determined from XPS data presented in Fig. 12 is the same in short and long synthesized samples and equal to ~529.6 eV. According to Zhang et al. [50] this peak corresponds to $O_2^-$ ions in the BFO lattice.

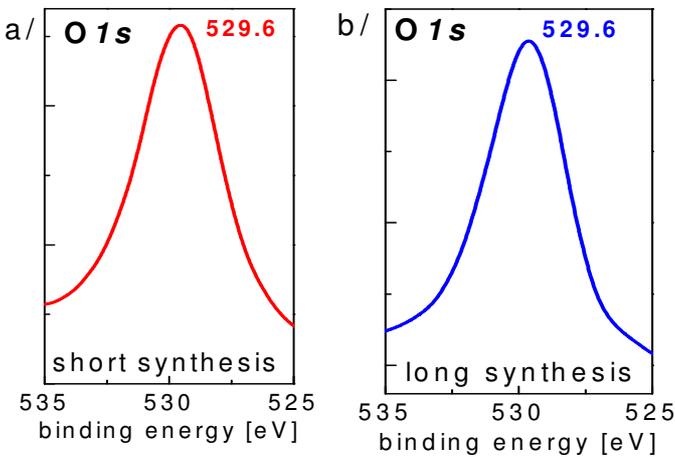

Figure 12. XPS signal from O*1s* core level signals for the sample synthesised during short time $t_s$=20 min a) and long time $t_s$=60 min b)

## IV. CONCLUSIONS

The BFO flowers synthesized by means of microwave assisted process consists mainly of BiFeO$_3$ and traces of Bi$_2$O$_3$ oxide and metallic Bi and Fe deposited mainly at the surface of theirs petals. The presence of impurity phases is more pronounced in the flowers synthesized during the short time $t_s$=20 min, and almost disappear, if the time of processing approaches $t_s$=60 min. For the short synthesis the content of Fe ions of various valence is comparable *i.e.* 51% of $Fe^{2+}$ and 49% of $Fe^{3+}$ ions. The nanoflowers synthesized during the long time exhibit unequal ratio of Fe ions with various valence; 58% of $Fe^{2+}$ and 42% of $Fe^{3+}$ ions. The samples obtained with the synthesis contain also large amounts of amorphous or nanocrystalline phase. The nanoflowers synthesized during long time are composed mainly of nanocrystallites. The petals of the nanoflowers, regardless their shape and thickness, are always composed of the same "building blocks" *i.e.* nanocrystals with the size about 35-39 nm. The thickness of the petals can be controlled by the amount of the surfactant added, and varies from 78 nm to about 420 nm. The growth and shape of the nanoflowers is determined by the process of diffusion in the initial stages of reaction resulting in petal-like morphology. On the other hand, the ball-like shape of the nanoflowers allows decreasing total surface energy of the BFO crystallites agglomeration. The BFO nanoflowers were found to exhibit enhanced magnetization [43]. Our XPS studies show that the effect may be not only related to the size effect but also to the $Fe^{2+}$ ions apparent at the petal surfaces. The $Fe^{2+}$ ions have the magnetic moment of ~ 6.7 $\mu_B$, which is higher in comparison with that of $Fe^{3+}$ ions equal to ~5.9 $\mu_B$.

## V. ACKNOWLEDGEMENTS

This project has been supported by National Science Centre (project No. N N507 229040). B.H. was supported by COST Action MP0904. M.M. was supported through the European Union - European Social Fund and Human Capital - National Cohesion Strategy. We would like to thanks to prof. B. Łęska and dr R. Pankiewicz from Adam Mickiewicz University for assistance in microwave hydrothermal method.

### REFERENCES

[1] N. Wintjes, J. Lobo-Checa, J. Hornung, T. Samuely, F. Diederich, T.A. Jung, Two-dimensional phase behavior of a biomolecular porphyrin system at the solid-vacuum interface. J Am Chem Soc 132 (2010) 7306–7311.

[2] S.N. Barnaby, N.H. Sarker, A. Tsiola, I.A. Banerjee, Biomimetic formation of chicoric-acid-directed luminescent silver nanodendrites. Nanotechnol 23 (2012) 294011.

[3] V. Mathur, J. Raj, K. Chouhan, V. Thanvi, Nokia morph technology. Int J Eng Res and Technol 2 (2013) 34-38.

[4] S.H. Ko, D. Lee, H.W. Kang, K.H. Nam, J.Y. Yeo, S.J. Hong, C.P. Grigoropoulos, H.J. Sunget, Nanoforest of hydrothermally grown hierarchical ZnO nanowires for a high efficiency dye-sensitized solar cell. Nano Lett 11 (2011) 666.

[5] B.I. Kharisov, A review for synthesis of nanoflowers. Rec Pat on Nanotechnol 2 (2008) 190-200.

[6] MRS website: http://www.mrs.org/science-as-art/

[7] A. Tiwari, M. Snure, Synthesis and characterization of ZnO nano-plant-like electrodes. J Nanosci Nanotechnol 8 (2008) 3981–3987.

[8] Y.B. Li, Y. Bando, D. Golberg, MoS2 nanoflowers and their field emission properties. Appl Phys Lett 82 (2003) 1962.

[9] Y. Ligang, Z. Gengmin, W. Yue, B. Xin, G. Dengzhu, Cupric oxide nanoflowers synthesized with a simple solution route and their field emission. J Cryst Growth 310 (2008) 3125-3130.

[10] Q. Yan, X. Li, Q. Zhao, G. Chen, Shape-controlled fabrication of the porous Co$_3$O$_4$ nanoflower clusters for efficient catalytic oxidation of gaseous toluene. J Hazard Mat 209–210 (2012) 385–391.

[11] Z. He, Q. Cai, H. Fang, G. Situ, J. Qiu, S. Song, J. Chen, Photocatalytic activity of TiO$_2$ containing anatase nanoparticles and rutile nanoflower structure consisting of nanorods. J Env Sci, 25 (2013) 2460-2468.

[12] X.-S. Fang, Ch.-H. Ye, T. Xie, Z.-Y. Wang, J.-W Zhao, L.-D. Zhang, Regular MgO nanoflowers and their enhanced dielectric responses. Appl Phys Lett 88 (2006) 013101.

[13] C. Li, G. Fang, N. Liu, Y. Ren, H. Huang, X. Zhao, Snowflake-like ZnO structures: Self-assembled growth and characterization. Mater Lett 62 (2008) 1761-1764.

[14] J. Sun, J. Ge,W. Liu, M. Lan, H. Zhang, P. Wang, Y. Wang, Z. Niu, Multi-enzyme co-embedded organic–inorganic hybrid nanoflowers: synthesis and application as a colorimetric sensor. Nanoscale 6 (2014) 255-262.

[15] J Xie, Q Zhang, J.Y Lee, D.I.C. Wang, The synthesis of SERS-active gold nanoflower tags for in vivo applications. ACS Nano 2 (2008) 2473-2480.

[16] T.T. Nhung. I.J. Kang, S.W. Lee, Fabrication and characterization of gold nanoflowers via chitosan-tripolyphosphate template films for biomedical applications. J Nanosci Nanotechnol 13 (2013) 5346-5350.

[17] G. Zhu R. Hu, Z. Zhao, Z, Chen, X, Zhang, W, Tan, Noncanonical Self-Assembly of Multifunctional DNA Nanoflowers for Biomedical Applications. J Am Chem Soc 135 (2013) 16438–16445.

[18] S. Kumari, R. P. Singh, Glycolic acid-g-chitosan-gold nanoflower nanocomposite scaffolds for drug delivery and tissue engineering. International J Biol Macromolecules, 50 (2012) 878–883.

[19] Y. Zhang, K. Yu, G. Li, D. Peng, Q. Zhang, F. Xu, W. Bai, S. Ouyang, Z. Zhu, Synthesis and field emission of patterned $SnO_2$ nanoflowers. Materials Letters 60 (2006) 3109–3112.

[20] N. Xiaomin, Z. Yongfeng, T. Dayong, Z. Huagui, W. Xingwei, Synthesis and characterization of hierarchical NiO nanoflowers with porous structure. J Cryst Growth 306 (2007) 418-421.

[21] P.B. Hui, X.L. Xu, M.Y. Guang, H.Y. Yun, Hydrogen peroxide biosensor based on electrodeposition of zinc oxide nanoflowers onto carbon nanotubes film electrode. Chinese Chem Lett 19 (2008) 314-318.

[22] Y. Gao, Z. Wang, S.X.Y. Liu, Y. Qian, Influence of anions on the morphology of nanophase alpha-$MnO_2$ crystal via hydrothermal process. J Nanosci Nanotech 6 (2006) 2576-2579.

[23] G. Li, L. Jiang, S. Pang, H. Peng, Z. Zhang, Molybdenum trioxide nanostructures: The evolution from helical nanosheets to crosslike nanoflowers to nanobelts. J Phys Chem B 110 (2006) 24472-24475.

[24] B.D. Liu, Y. Bando, C.C. Tang, D. Golberg, R.G. Xie, T. Sekiguchi, Synthesis and optical study of crystalline GaP nanoflowers. Appl Phys Lett 86 (2005) 083107.

[25] J. Du, Z. Liu, Z. Li, B. Han, Z. Sun, Y. Huang, Carbon nanoflowers synthesized by a reduction-pyrolysis-catalysis route. Mater Lett 59 (2005) 456-458.

[26] K. Das, A. Datta, S. Chaudhuri, $CuInS_2$ flower vaselike nanostructure arrays on a Cu Tape substrate by the copper indium sulfide on Cu-Tape (CISCuT) method: growth and characterization. Cryst Growth Des 7 (2007) 1547-1552.

[27] T.-T. Kang, X. Liu, R.Q. Zhang, W.G. Hu, G. Cong, F.-A. Zhao, Q. Zhu, InN nanoflowers grown by metal organic chemical vapor deposition. Appl Phys Lett 89 (2006) 071113.

[28] X. Cao, Q. Lu, X. Xu, J. Yan, H. Zeng, Single-crystal snowflake of $Cu_7S_4$: Low temperature, large scale synthesis and growth mechanism. Mater Lett 62 (2008) 2567-2570.

[29] L. Dong, Y. Chu, Y. Liu, M. Li, F. Yang, L. Li, Surfactant-assisted fabrication PbS nanorods, nanobelts, nanovelvet-flowers and dendritic nanostructures at lower temperature in aqueous solution. J Colloid Interface Sci 301 (2006) 503-510.

[30] W. Xu, K.Y. Liew, H. Liu,, T. Huang, C. Sun, Y. Zhao, Microwave assisted synthesis of nickel nanoparticles. Mater Lett 62 (2008) 2571-2573.

[31] I. Seung, C.B. Cha, K.T. Mo, S.H.H. Kim, Ferromagnetic cobalt nanodots, nanorices, nanowires and nanoflowers by polyol process. MRS Website 20 (2007) 2148-2153.

[32] M.-T. Hsiao, S.-F. Chen, D.-B. Shieh, C.-S. Yeh, One-pot synthesis of hollow $Au_3Cu$ spherical-like and biomineral botallackite $Cu_2(OH)_3Cl$ flowerlike architectures exhibiting antimicrobial activity. J Phys Chem B 110 (2006) 205-210.

[33] D.P. Amalnerkar, H.-Y. Lee, Y.K. Hwang, D.-P. Kim, J.-S. Chang, Swift Morphosynthesis of hierarchical nanostructures of CdS via microwave-induced semisolvothermal route. J Nanosci Nanotech 7 (2007) 4412-4420.

[34] S.H. Lee, Y.J. Kim, J. Park, Shape evolution of ZnTe nanocrystals: nanoflowers, nanodots, and nanorods. Chem Mat 19 (2007) 4670-4675.

[35] T.T. Kang, A. Hashimoto, A. Yamamoto, Optical properties of InN containing metallic indium. Appl Phys Lett 92 (2008) 111902.

[36] J. Ge, J. Lei, R.N. Zare, Protein-inorganic hybrid nanoflowers. Nat Nanotechnol 3 (2012) 428-32.

[37] M. Sakar, S. Balakumar, P. Saravanan, S.N. Jaisankar, Annealing temperature mediated physical properties of bismuth ferrite ($BiFeO_3$) nanostructures synthesized by a novel wet chemical method. Mat Res Bull 48 (2013) 2878-2885.

[38] B. Andrzejewski, K. Chybczyńska, K. Pogorzelec-Glaser, B. Hilczer, T. Toliński, B. Łęska, R. Pankiewicz, P. Cieluch, Magnetic relaxation in bismuth ferrite micro-cubes. Ferroelectrics 448 (2013) 58-70.

[39] D. Khomskii, Classifying multiferroics: Mechanisms and effects. Physics 2 (2009) 20.

[40] A.M. Kadomtseva, Yu.F. Popov, A.P. Pyatakov, G.P. Vorobev, A.K. Zvezdin, D. Viehland, Phase transitions in multiferroic $BiFeO_3$ crystals, thin-layers and ceramics: enduring potential for a single phase, room-temperature magnetoelectric "holy grail". Phase Trans 79 (2006) 1019-1042.

[41] I. Sosnowska, T.P. Neumaier, E. Steichele, Spiral magnetic ordering in bismuth ferrite. J Phys C: Solid State Phys 15 (1982) 4835-4846.

[42] B. Andrzejewski, A. Molak, B. Hilczer, A. Budziak, R. Bujakiewicz-Korońska, Field induced changes in cycloidal spin ordering and coincidence between magnetic and electric anomalies in $BiFeO_3$ multiferroic. Journal of Magnetism and Magnetic Materials 342 (2013) 17-26.

[43] K. Chybczyńska, P. Ławniczak, B. Hilczer, B. Łęska, R. Pankiewicz, A. Pietraszko, L. Kępiński, T. Kałuski, P. Cieluch, F. Matelski; Synthesis and properties of bismuth ferrite multiferroic flowers. J Mater Sci 49 (2014) 2596-2604.

[44] G. Catalan, J.F. Scott, Physics and applications of bismuth ferrite. Adv Mat 21 (2009) 2463-2485.

[45] S.-J.L. Kang, Sintering densification, grain growth & microstructure. Elsevier, Oxford 2005.

[46] G. Wulf, Zur Frage der Geschwindigkeit des Wachstums und der Auflösung von Kristallflächen. Z Kristallogr 34 (1901) 449.

[47] C. Herring, Some theorems on the free energies of crystal surfaces. Phys Rev 82 (1951) 87.

[48] J.C. Denardin, A.L. Brandl, M. Knobel, P. Panissod, A.B. Pakhomov, H. Liu, X.X. Zhang, Thermoremanence and zero-field-cooled/field-cooled magnetization study of $Co_x(SiO_2)_{1-x}$ granular films. Phys Rev B 65 (2002) 064422-1-8.

[49] P. Scherrer, Bestimmung der Grösse und der inner Structur von Kolloidteilchen mittels Röntgenstrahlen. Nachrichten Gesell Wiss Göttingen 26 (1918) 98–100.

[50] S.T. Zhang, M.H. Lu, D. Wu, Y.F. Chen, N.B. Ming, Larger polarization and weak ferromagnetism in quenched $BiFeO_3$ ceramics with a distorted rhombohedral crystal structure. Appl Phys Lett 87 (2005) 262907.